# GEOMETRICAL-INDUCED RECTIFICATION IN TWO-DIMENSIONAL BALLISTIC NANODEVICES

**Daniela Dragoman[1(a)] and Mircea Dragoman[2]**

[1]Univ. Bucharest, Physics Dept., P.O. Box MG-11, 077125 Bucharest, Romania

[2]National Research and Development Institute in Microtechnologies, Str. Erou Iancu Nicolae 126 A , 077190 Bucharest, Romania

The paper demonstrates that a two-dimensional ballistic nanodevice in which the electron gas satisfies either the Schrödinger equation (as in quantum wells in common semiconductor heterostructures) or the Dirac equation (as in graphene) is able to rectify an electric signal if the device has a non-uniform cross section, for instance a taper configuration. No *p-n* junctions or dissimilar electrodes are necessary for rectification.

(a) Corresponding author email: danieladragoman@yahoo.com

## 1. Introduction

The rectification process transforms an alternative signal into a DC signal. Rectification is a key signal processing tool in many applications, ranging from power supplies up to high-frequency detectors, and is present nowadays in smart phones, computers, TV-sets and radios wireless communications. Rectification is implemented using a semiconducting *p-n* junction obtained by chemical doping. However, in two-dimensional electron gas (2DEG) devices based on semiconducting heterostructures, carbon nanotubes or graphene, chemical doping becomes difficult and is replaced by electrostatic doping performed by multiple metallic electrodes polarized with different DC voltages [1-3]. Also, in CNTs rectification can be achieved in Schottky contacts implemented with the help of dissimilar or asymmetric electrodes made from different metals [4,5]. However, metallic gates on nanostructures complicate very much the architecture of rectifying devices, use extensive nanolithography techniques, and quite often introduce parasitic effects in the gated rectifying nanodevices.

Therefore, it is very important to be able to rectify alternative signals based on a different mechanism than chemical doping or Schottky contacts. Such a different mechanism is the geometrical asymmetry in 2DEG nanodevices with non-uniform cross-sections, for example, in taper configurations. Tailoring the shape of semiconductor channels was used already in self-switching devices, but was not related to ballistic transport [6], while a unipolar diode that acts as room-temperature THz detector based on symmetry breaking of a semiconducting channel was demonstrated recently [7]. Geometrical asymmetries are used for metal-insulator-metal diodes working in a non-ballistic regime to enhance the sensitivity of electromagnetic wave detection at 900 MHz, which remains, however, quite low [8]. Recently, a graphene trapezoidal-shaped diode deposited on $Si/SiO_2$ and coupled to a bowtie metallic antenna was shown to detect an electromagnetic radiation at 28 THz generated by a $CO_2$ laser [9]. However, the rectification properties of this diode are very poor and its nonlinearities are vey weak even at gate voltages of 40 V. These poor results are caused by the high losses of the doped Si substrate, which appear even at few GHz [10], and by the non-ballistic transport of charge carriers. At room temperature, the ballistic transport in graphene occurs for a mean-free path of 0.4 μm, reaching an intrinsic mobility of 44 000 cm$^2$V$^{-1}$s$^{-1}$ [11], mean-free carrier paths longer than 1 μm and mobilities higher than 100 000 cm$^2$/s being observed only in the exceptional case when graphene is deposited on a hexagonal boron nitride substrate [12] matching the graphene lattice.

So, the role of this paper is to investigate the rectifying properties of a 2DEG nanodevice with a tapered geometry working in the ballistic regime, where the transport of charge carriers is described by either the Schrödinger or the Dirac equation. We show that such a device can rectify alternative electrical signals under certain conditions, at room temperature, thus paving the way for generation of THz signals via multiplication and for detection of THz signals used in imaging applications.

## 2. Signal rectification of a tapered 2DEG satisfying the Schrödinger equation

A 2DEG satisfying the Schrödinger equation is in fact a quantum well formed in a common semiconductor heterostructure. A schematic representation of a tapered 2DEG connected at two metallic electrodes is displayed in Fig. 1; we assume throughout this paper that $d_{in} > d_{out}$. Rectification occurs for current flowing along the $x$ direction. We assume that the boundaries of the tapered 2DEG act as regions with an infinite energy potential and model the transport of charge carriers by replacing the taper with a number of $N$ regions with constant but different widths shown as rectangles with dotted boundaries in Fig. 1. In each region of width $d_j$, $j = in,1,\ldots,N,out$, the wavevector component along $y$ has discrete values $n\pi/d_j$, so that the wavevector component along $x$ is equal to $k_{n,j} = \sqrt{2m(E-V_j)/\hbar^2 - (n\pi/d_j)^2}$, where $m$ and $E$ are the effective mass and energy of the electron, respectively, and $V_j$ is the potential energy in region $j$. A solution of the Schrödinger equation corresponding to a given $n$ value for which $k_{n,j}$ is real is referred to as a mode (no charge transport takes place for an imaginary $k_{n,j}$ since tunneling place no role in this device); the number of modes in the $j$th region is $N_j = \text{Int}[d_j\sqrt{2m(E-V_j)}/\pi\hbar]$, where $\text{Int}[]$ denotes the integer part and $d_j = d_{in} - (d_{in} - d_{out})j/(N+1)$, for $j = 1,\ldots,N$. We assume throughout this paper that a bias voltage $V$ is applied across the tapered structure, such that $V_{in} = 0$, $V_{out} = -eV$, and $V_j = -jeV/(N+1)$, $j = 1\ldots N$.

The rectification principle can be understood from Fig. 2, which shows the variation of the ratio of outgoing and incoming mode numbers with $V$ for three energy values: $E = 0.1$ eV (green dashed line), 0.2 eV (solid black line) and 0.3 eV (red dotted line). We consider throughout this paper $d_{in} = 50$ nm, $d_{out} = 10$ nm and $L = 100$ nm. Figure 2 shows that the number of emerging modes, and hence the current, which is proportional to the mode number in the Landauer theory, depends on the applied voltage: the current is high for one polarity and decreases to zero for the other polarity, as the number of outgoing modes decreases.

A more refined analysis of charge transport in the tapered 2DEG must take into account that scattering occurs between different modes in quantum waveguides with a non-uniform cross-section. Therefore, the transmission probability is determined by the requirement of flux conservation and is given by

$$T = \sum_{n=1}^{N_{out}} k_{n,out} \mid A_{n,out}\mid^2 \Big/ \sum_{n=1}^{N_{in}} k_{n,in} \mid A_{n,in}\mid^2 \qquad (1)$$

where the coefficients $A_{n,j}$, $j = in, out$ represent the amplitudes of the forward-propagating components of the wavefunction in the incidence and outgoing media, respectively, and are found by imposing the continuity conditions for the wavefunction

$$\Psi_j(x,y) = \sum_{n=1}^{N_j}[A_{n,j}\exp(ik_{n,j}x) + B_{n,j}\exp(-ik_{n,j}x)]\sin(2n\pi y/d_j) \qquad (2)$$

and its derivative with respect to $x$ at each interface between different regions (for details on electron propagation in waveguides with non-uniform cross-section, see [13,14]).

The voltage dependence of the transmission coefficient for $E = 0.2$ eV is represented in Fig. 3 for the case of a step interface between the incoming and outgoing media, i.e. for $N = 0$ (red dotted line), and for $N = 1$ (solid black line) and 2 (green dashed line). As $N$ increases, the transmission coefficient resembles the smooth curve represented with magenta dashed-dotted line in Fig. 3. As the number of modes, the transmission through the tapered configuration depends asymmetrically on the applied voltage: it takes high values for positive biases and low values for negative biases, indicating no current transport for $V$ values for which there are no propagating outgoing modes. As suggested by Fig. 2, the bias voltage for which $T$ increases sharply increases as the energy decreases.

The current, calculated with the Landauer formula, is represented in Fig. 4 for three values of the Fermi energy level: $E_F = 0$ (green dashed line), 0.1 eV (solid black line) and 0.2 eV (red dotted line). Although current rectification occurs in all cases, the shape of the $I$-$V$ characteristic depends on the position of the Fermi level. A low $E_F$ corresponds to lower energy values of the charge carrying electrons, for which $T$ can have significant values for only one polarity (see the green dashed line), while for high enough Fermi levels the transmission probability has a finite value at $V = 0$ (see also Fig. 3), so that current flows also for a limited range of negative $V$ values. Therefore, to have current flow only for one polarization, one must choose $E_F$ appropriately.

**3. Signal rectification of a tapered graphene sheet**

Graphene is a natural 2DEG, in which the charge carriers satisfy a massless Dirac equation. As a result, an infinite graphene sheet has no bandgap, although a finite bandgap exists in narrow graphene strips, called also nanoribbons. In our tapered configuration, there is a small bandgap because the spinorial wavefunction cannot extend outside the taper, and hence such a taper could rectify an electrical signal. However, unlike for Schrödinger-type charge carriers, rectification is possible only for electrical signals with small enough amplitudes.

The mathematical treatment of the graphene taper parallels that of the previous Section. In particular, the taper is again modeled as a succession of $N$ regions with widths $d_j$, $j = in,1,\ldots,N,out$, so that the wavevector component along $y$ is again discrete and given by the same expression as in the previous section, the wavenumbers along $x$ being now $k_{n,j} = \mathrm{sgn}(E - V_j)\sqrt{(E - V_j)^2 /(\hbar^2 v_F^2) - (n\pi / d_j)^2}$ [15], where $v_F \cong c / 300$ is the Fermi velocity. If $\mathrm{sgn}(E - V_j)$ is positive the electrical charge is carried by electrons, otherwise by holes. In the tapered waveguide charge propagation occurs for real $k_{n,j}$ values, the number of modes in the $j$th region being now given by $N_j = \mathrm{Int}[d_j \mid E - V_j \mid /(\pi \hbar v_F)]$. The expressions of $d_j$ and $V_j$ are the same as in the previous case.

In the graphene taper, with identical dimensions as those considered in Section 2, the voltage dependence of the ratio of outgoing and incoming mode numbers is represented in Fig. 5 for $E = 0.1$ eV (green dashed line), 0.2 eV (solid black line) and 0.3 eV (red dotted line). Although there are a finite number of outgoing modes for both voltage polarizations, current rectification can be achieved if we exploit the small voltage range in which $N_{out}$, and hence $T$ vanishes. The width of this voltage range is given by $\pi \hbar v_F / d_{out}$: it increases as $d_{out}$ decreases. Please note that this voltage range does not depend on the energy of the incident electrons, but shifts to lower values as the energy increases.

The transmission coefficient is now defined as

$$T = \sum_{n=1}^{N_{out}} |A_{n,out}|^2 \Big/ \sum_{n=1}^{N_{in}} |A_{n,in}|^2 \qquad (3)$$

where the coefficients where the coefficients $A_{n,j}$, $j = in, out$ are calculate by imposing the continuity condition at each interface between different regions for the two components of the spinorial wavefunction

$$\Psi_j(x,y) = \begin{pmatrix} \sum_{n=1}^{N_j} [A_{n,j} \exp(ik_{n,j}x) + B_{n,j} \exp(-ik_{n,j}x)] \sin(2n\pi y / d_j) \\ \sum_{n=1}^{N_j} [A_{n,j} \exp(ik_{n,j}x) - B_{n,j} \exp(-ik_{n,j}x)] \sin(2n\pi y / d_j) \end{pmatrix}. \qquad (4)$$

Figure 6 represent the transmission coefficient dependence on $V$ for $E = 0.2$ eV and for $N = 0$ (red dotted line), 1 (solid black line) and 2 (green dashed line). $T$ tends to the magenta dashed-dotted line for large $N$ values. Unlike for Schrödinger-type charge carriers, the transmission has significant values, except for a narrow voltage range in which there are no outgoing modes.

The current through the tapered graphene, calculated also with the Landauer formula, is represented in Fig. 7 for three values of the Fermi energy level: $E_F = 0$ (green dashed line), 0.1 eV (solid black line) and 0.2 eV (red dotted line). As follows from Fig. 7, the shape of the $I$-$V$ characteristics is the same in all cases, but the current amplitude depends on the Fermi level (on the energy of electrons that contribute to the current flow). Unlike for Schrödinger-type charge carriers, the $I$-$V$ characteristics are symmetric around a bias voltage that represents the center of the region in which the transmission coefficient/number of outgoing modes vanishes. Current rectification can occur only if the Fermi level is chosen such that the sudden current increase or decrease takes place at $V = 0$. Such a situation is represented with the red dotted line in Fig. 7. In this case current rectification, in the sense that finite current values exist for one polarization only, occurs only if the incident voltage signal has an amplitude (in this simulation) lower than 0.4 V. Voltage signals with higher amplitudes could be rectified if $d_{out}$ decreases.

**Conclusions**

We have demonstrated that rectification, which is a key electronic function, takes place in two-dimensional ballistic nanodevices, in which the electron gas is described by either the Schrödinger equation or the Dirac equation, only when the 2DEG geometry is asymmetric with respect to current flow direction. While in quantum wells described by Schrödinger equation the rectification is produced for a single, positive polarity of the electrical field (or negative polarity in two-dimensional hole gases), in the case of graphene, which satisfy a Dirac-like equation, rectification is ambipolar; ambipolarity is an imprint of all graphene device. Because the cutoff frequencies of the above devices is located in the THz region [16], our results open up the perspective of generating THz frequencies using multipliers with a much simpler architecture and of imaging THz radiation by connecting the above rectifiers with small antennas on a single Si wafer.

**References**

[1]  D.J. Reilly, and C.M. Marcus, "Fast single-charge sensing with a rf quantum point contact", Appl. Phys. Lett. 91, 162101 (2007).

[2]  J.U. Lee, P.P. Gipp, and C.M. Heller, "Carbon-nanotube *p-n* junction diodes", Appl. Phys. Lett. 85, 145 (2004).

[3]  J.R. Williams, L. Dicarlo and C.M. Marcus, "Quantum-controlled *p-n* junctions of graphene", Science 317, 638 (2007).

[4]  M.H. Yang, K.B.K. Teo, and W.I. Milne, "Carbon nanotube Schottky diode and directionally depend field-effect transistor using asymmetric contacts", Appl. Phys. Lett. 87, 253116 (2005).

[5]  C. Lu, L. An, Q. Fu and J. Liu, "Schottky diodes from asymmetric metal-nanotube contacts", Appl. Phys. Lett. 88, 133501 (2006).

[6]  A.M. Song, M. Missous, P. Omling, A.R. Peaker, L. Samuelson, and W. Seifert, "Unidirectional electron flow in a nanometer-scale semiconductor channel: A self-switching device", Appl. Phys. Lett. 83, 1881 (2003).

[7]  C. Balocco, S.R. Kasjoo, X.F. Lu, L.Q. Zhang, Y. Alimi, S. Winnerl, and A.M. Song, "Room temperature of a unipolar nanodiode at terahertz frequencies", Appl. Phys. Lett. 98, 223501 (2011).

[8]  K. Choi, G. Ryu, F. Yesilkoy, A. Chryssis, N. Goldsman, M. Dagenais, and M. Peckerar, "Geometry enhanced asymmetric rectifying tunnel diodes", J. Vac. Sci B28, C6050 (2010).

[9]  G. Moddel, Z. Zhu, S. Grover, and S. Joshi, "Ultrahigh speed graphene diode with reversible polarity", Solid Sate Commun., doi:10.1016/j.ssc.2012.06.013 (2012).

[10] G. Deligiorgis, M. Dragoman, D. Neculoiu, D. Dragoman, G. Konstantinidis, A. Cismaru, and R. Plana, "Microwave propagation in graphene", Appl. Phys. Lett. 95, 073107 (2009).

[11] R.S. Shishir and D.K. Ferry, "Intrinsic mobility in graphene", J. Phys.: Condens. Matter. 21, 23204 (2009).

[12] A.S. Mayorov, R.V. Gorbachev, S.V. Morozov, L. Britnell, R. Jail, L.A. Ponomarenko, K.S. Novoselov, K. Watanabe, T. Taniguchi, and A.K. Geim, "Micrometer-scale ballistic transport in encapsulated graphene at room temperature", Nano Lett. 11, 2396 (2011).

[13] H. Wu, D.W.L. Sprung, J. Martorell, and S. Klarsfeld, "Quantum wire with periodic serial structure", Phys. Rev. B 44, 6351 (1991).

[14] A. Weisshaar, J. Lary, S.M. Goodnick, and V.K. Tripathi, "Analysis of discontinuities in quantum waveguide structures", Appl. Phys. Lett. 55, 2114 (1989).

[15] D. Dragoman, "Evidence against Klein paradox in graphene", Phys. Scr. 79, 015003, (2009).

[16] D. Dragoman and M. Dragoman, "Time flow in graphene and its implications on cutoff frequency of ballistic graphene devices", J. Appl. Phys. 110, 014302 (2011).

**Figure Captions**

Fig. 1 Schematic representation of the ballistic rectifier.

Fig. 2 Voltage dependence of the ratio of outgoing and incoming mode numbers in a 2DEG for $E = 0.1$ eV (green dashed line), 0.2 eV (solid black line) and 0.3 eV (red dotted line).

Fig. 3 Voltage dependence of the transmission coefficient in a 2DEG for $E = 0.2$ eV and $N = 0$ (red dotted line), 1 (solid black line) and 2 (green dashed line). For large $N$, $T$ tends to the magenta dashed-dotted line.

Fig. 4 Current-voltage characteristics in a 2DEG for $E_F = 0$ (green dashed line), 0.1 eV (solid black line) and 0.2 eV (red dotted line).

Fig. 5 Voltage dependence of the ratio of outgoing and incoming mode numbers in graphene for $E = 0.1$ eV (green dashed line), 0.2 eV (solid black line) and 0.3 eV (red dotted line).

Fig. 6 Voltage dependence of the transmission coefficient in graphene for $E = 0.2$ eV and for $N = 0$ (red dotted line), 1 (solid black line) and 2 (green dashed line). $T$ tends to the magenta dashed-dotted line for large $N$ values.

Fig. 7 Current-voltage characteristics in graphene for $E_F = 0$ (green dashed line), 0.1 eV (solid black line) and 0.2 eV (red dotted line).

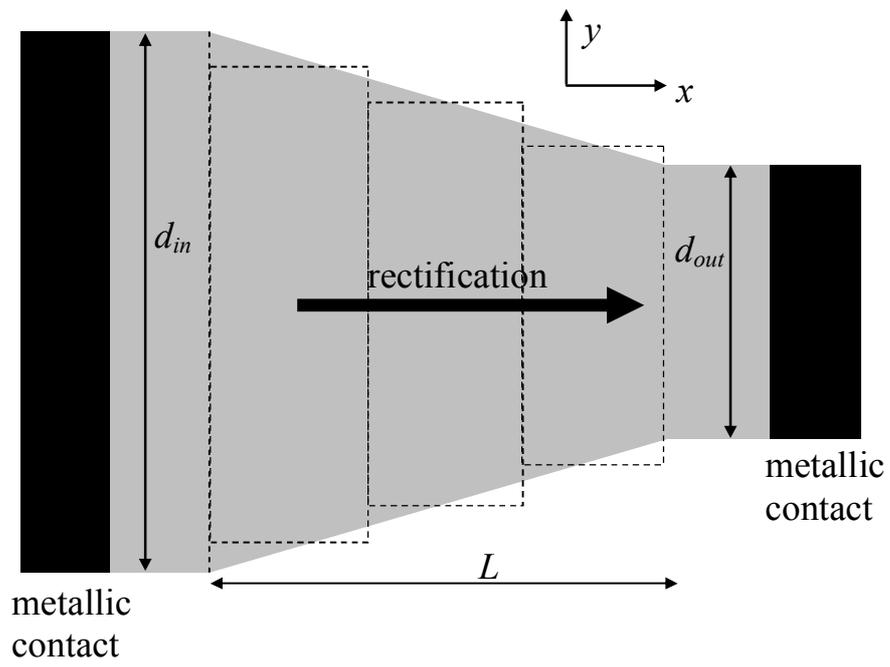

Fig. 1

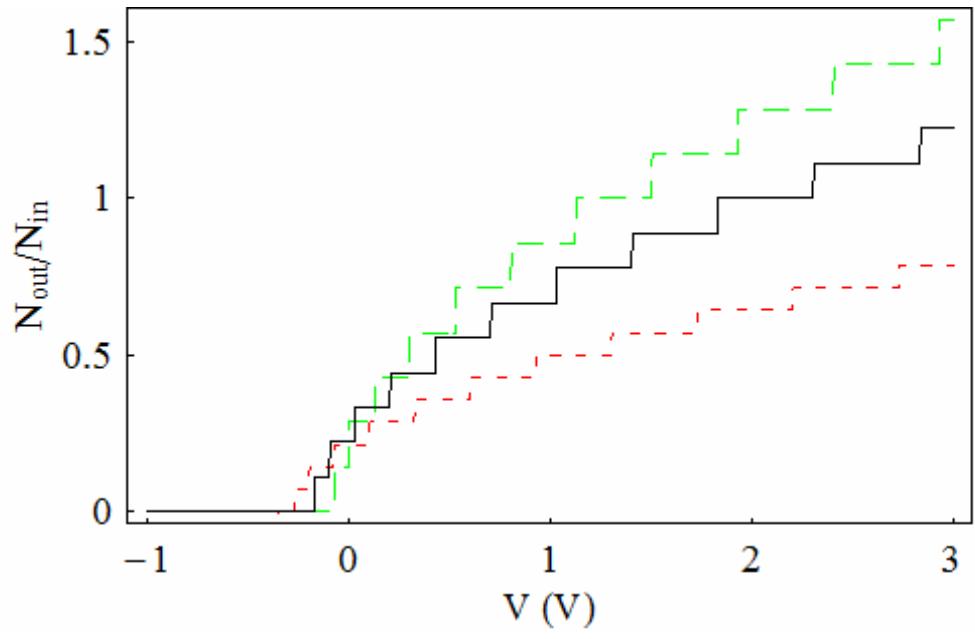

Fig. 2

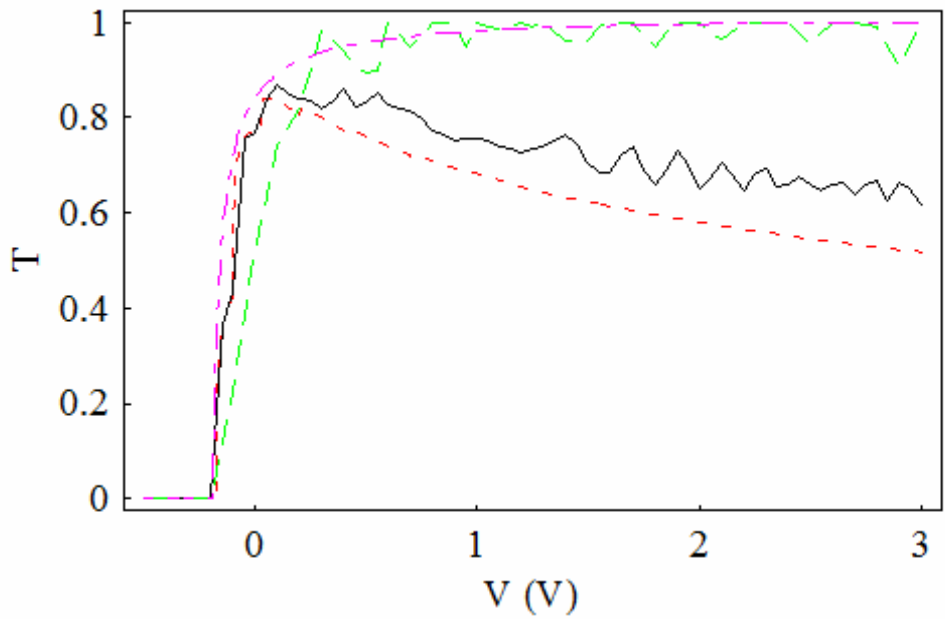

Fig. 3

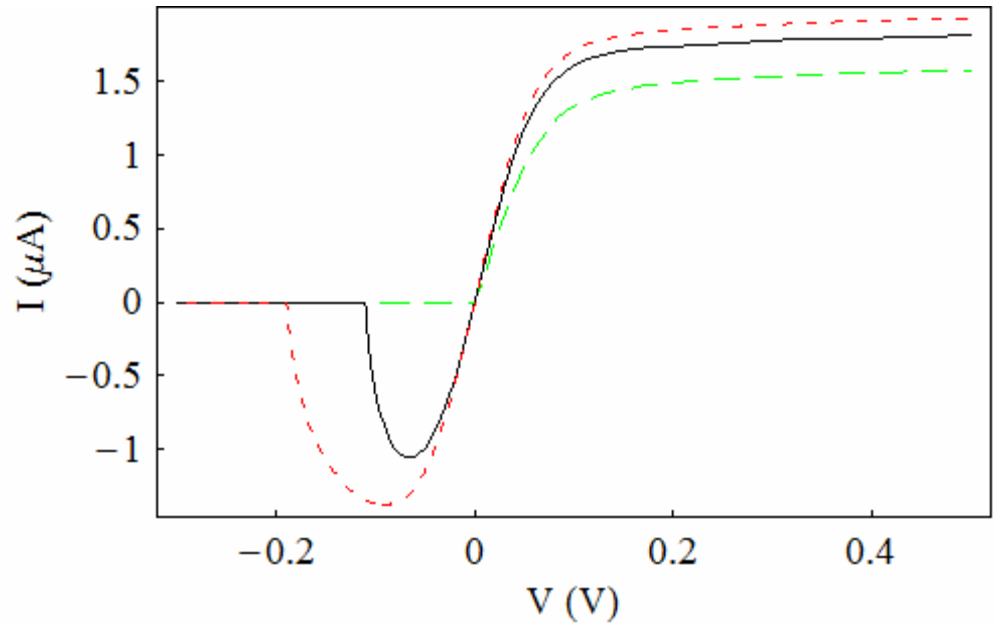

Fig. 4

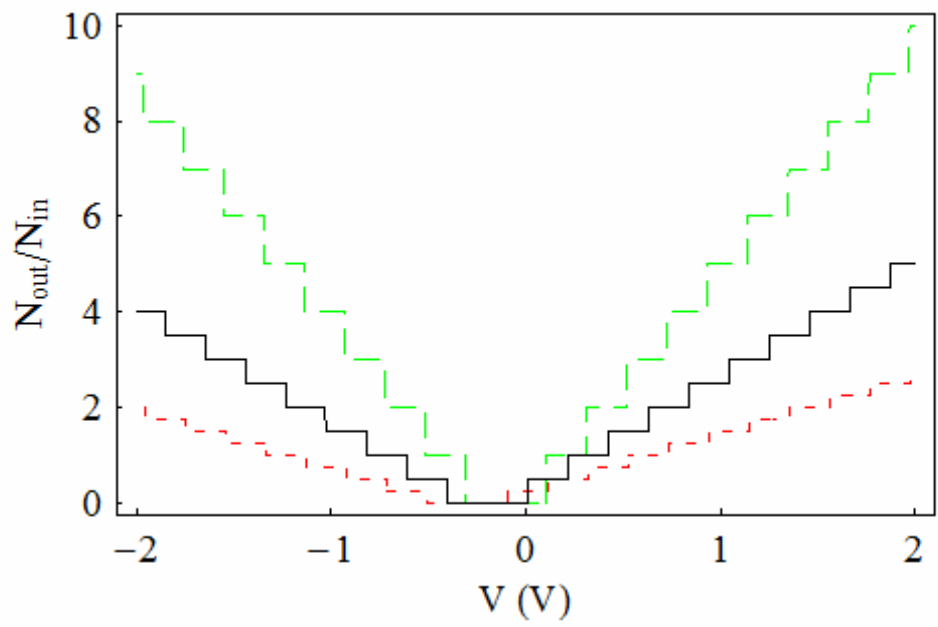

Fig. 5

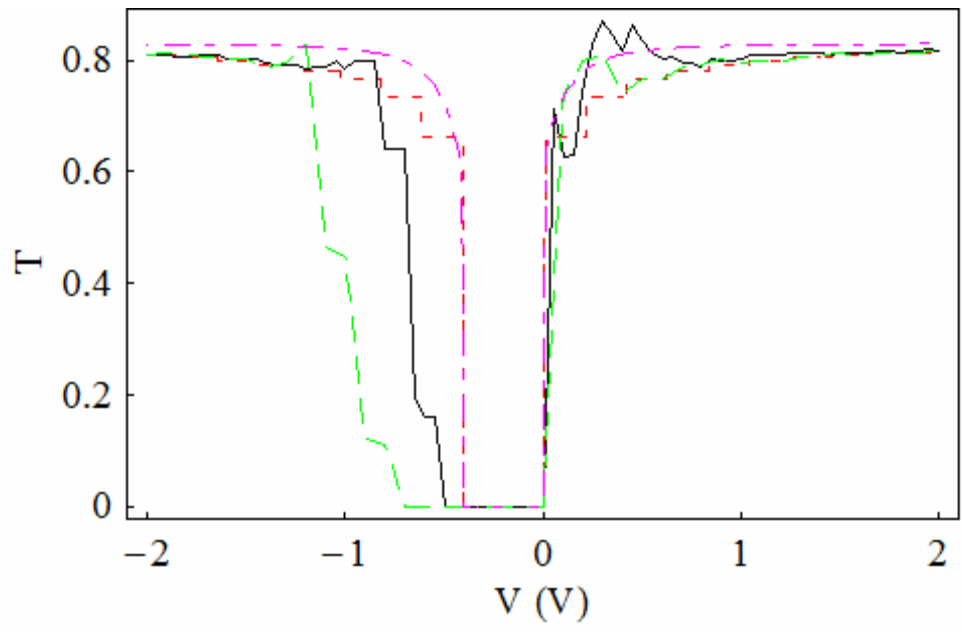

Fig. 6

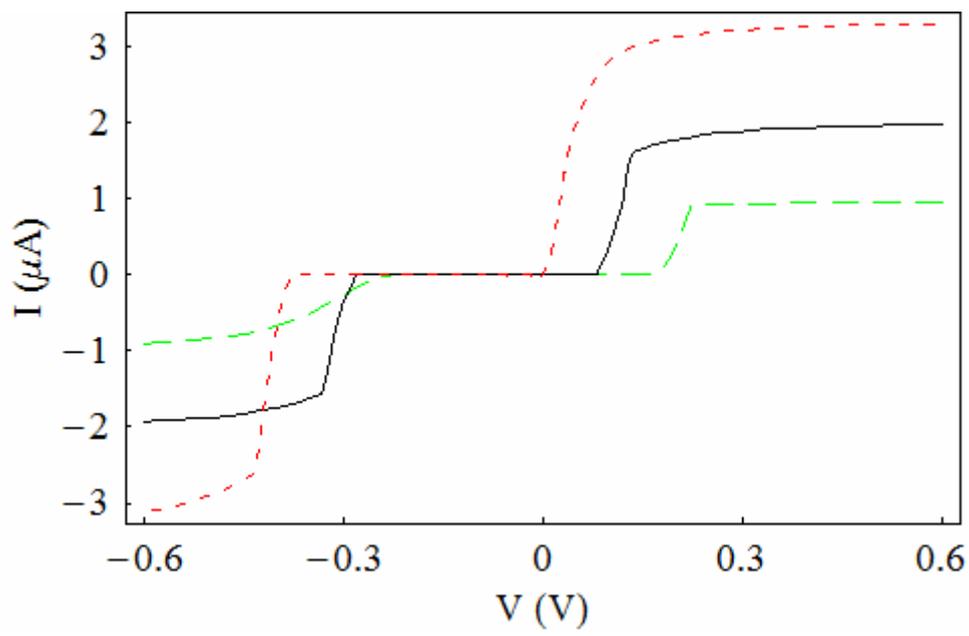

Fig. 7